\documentclass[oribibl]{llncs}

 \usepackage{makecell}
\usepackage[hidelinks]{hyperref}
\usepackage{graphicx}
\usepackage{algorithm}
\usepackage{algpseudocode}
\usepackage{url}
\usepackage{ulem}
\usepackage{mathtools}
\usepackage{scalerel}
\usepackage{setspace}
\usepackage[strict]{changepage}
\usepackage{caption}
\usepackage[letterspace=-50]{microtype}
\usepackage{afterpage}
\usepackage{ragged2e}
\usepackage{amsopn}
\usepackage{enumerate}
\usepackage{tikz}
\usetikzlibrary{trees}
\usepackage[utf8]{inputenc}
\usepackage{amsmath}
\usepackage{amssymb,amsfonts}
\usepackage[T1]{fontenc}
\usepackage{graphicx}
\usepackage{hyperref}
\usepackage{wrapfig}
\usepackage{array}
\usepackage{thmtools, thm-restate}
\usepackage{enumitem}

\title{On the power of counting the total number of computation paths of NPTMs}
\author{Eleni Bakali\inst{1} \and Aggeliki Chalki\inst{2} \and Sotiris Kanellopoulos\inst{1} \and Aris Pagourtzis\inst{1} \and Stathis Zachos\inst{1}}

\institute{School of Electrical and Computer Engineering, National Technical University of Athens, Greece, \email{mpakali@corelab.ntua.gr, sotkanellopoulos@mail.ntua.gr, pagour@cs.ntua.gr, zachos@cs.ntua.gr}
\and Department of Computer Science, Reykjavik University, Iceland, \email{angelikic@ru.is}
}
 
\renewenvironment{proof}{\noindent\textit{Proof}. }{\hfill $\Box$}

\renewenvironment{altproof}{\noindent\textit{Alternative proof}. }{\hfill $\Box$}

\newcommand{\totp}{\ensuremath{\mathsf{TotP}}}
\newcommand{\sP}{\ensuremath{\#\mathsf{P}}}

\newcommand{\cP}{\ensuremath{\mathsf{P}}}

\newcommand{\NP}{\ensuremath{\mathsf{NP}}}
\newcommand{\RP}{\ensuremath{\mathsf{RP}}}

\newcommand{\FP}{\ensuremath{\mathsf{FP}}}

\newcommand{\BPP}{\ensuremath{\mathsf{BPP}}}
\newcommand{\otp}{\ensuremath{\mathsf{\oplus_{tot} P}}}
\newcommand{\op}{\ensuremath{\mathsf{\oplus P}}}

\newcommand{\up}{\ensuremath{\mathsf{UP}}}
\newcommand{\utp}{\ensuremath{\mathsf{U_{tot}P}}}
\newcommand{\gaptp}{\ensuremath{\mathsf{Gap_{tot} P}}}
\newcommand{\gapp}{\ensuremath{\mathsf{GapP}}}
\newcommand{\pp}{\ensuremath{\mathsf{PP}}}
\newcommand{\ppt}{\ensuremath{\mathsf{P_{tot}P}}}
\newcommand{\cep}{\ensuremath{\mathsf{C_=P}}}
\newcommand{\cetp}{\ensuremath{\mathsf{{{C_=}_{tot}P}}}}
\newcommand{\modkp}{\ensuremath{\mathsf{Mod_k P}}}
\newcommand{\modktp}{\ensuremath{\mathsf{ {Mod_k}_{tot} P}}}
\newcommand{\fewp}{\ensuremath{\mathsf{FewP}}}
\newcommand{\fewtp}{\ensuremath{\mathsf{Few_{tot}P}}}
\newcommand{\spp}{\ensuremath{\mathsf{SPP}}}
\newcommand{\sppt}{\ensuremath{\mathsf{SP_{tot}P}}}
\newcommand{\wpp}{\ensuremath{\mathsf{WPP}}}
\newcommand{\wppt}{\ensuremath{\mathsf{WP_{tot}P}}}
\newcommand{\PH}{\ensuremath{\mathsf{PH}}}

\newcommand{\perfmatch}{\textsc{\#PerfMatch}}

\newcommand{\sSAT}{\textsc{\#Sat}}
\newcommand{\SAT}{\textsc{Sat}}

\newcommand{\dnfsat}{\textsc{\#DNF-Sat}}
\newcommand{\sizeofsubtree}{\textsc{Size-of-Subtree}}

\begin{document}

\maketitle
\begin{abstract}
        In this paper, we define and study variants of several complexity classes of decision problems that are defined via some criteria on the number of accepting paths of an NPTM. In these
        variants, we modify the acceptance criteria so that they concern the total number of computation paths,
        instead of the number of accepting ones. This direction reflects the relationship between the counting classes \sP{} and \totp{}, which are the classes of functions that count the number of accepting paths and the total number of paths of NPTMs, respectively. The former is the well-studied class of counting versions of \NP{} problems, introduced by Valiant (1979). The latter contains all self-reducible counting problems in \sP{} whose decision version is in \cP, among them prominent \sP-complete problems such as \textsc{Non-negative Permanent}, \perfmatch\ and \dnfsat, thus playing a significant role in the study of approximable counting problems. \smallskip
    
    We show that almost all classes introduced in this work coincide with their `\# accepting paths'-definable counterparts, thus providing an alternative model of computation for
    the classes \op, \modkp, \spp, \wpp, \cep, and \pp{}. Moreover, for each of these classes, 
    we present a novel family of complete problems which are defined via problems that are \totp-complete under parsimonious reductions. This way, we show that all the aforementioned classes have complete problems that are defined via counting problems whose existence version is in $\cP$, in contrast to the standard way of obtaining completeness results via counting versions of \NP-complete problems. To the best of our knowledge, prior to this work, such results were known only for \op\ and \cep. \smallskip
    
     We also build upon a result by Curticapean, to exhibit yet another way to obtain complete problems for \wpp\ and \pp, namely via the difference of values of the \totp\ function \perfmatch\ on pairs of graphs. Finally, for the so defined \wpp-complete problem, we provide an exponential lower bound under the randomized Exponential Time Hypothesis, showcasing the hardness of the class.
\end{abstract}


\section{Introduction}\label{intro}

Valiant introduced the complexity class \sP\ in his seminal paper~\cite{Valiant79} to characterize the complexity of the permanent function. \sP{} contains the counting versions of \NP\ problems and equivalently, functions that count the accepting paths of non-deterministic polynomial-time Turing machines (NPTMs). For example, \sSAT, i.e. the problem of counting the number of satisfying assignments of a propositional formula, lies in  \sP{}.  
The class of functions that count the total number of paths of NPTMs, namely \totp, was introduced and studied in \cite{KPSZ01,PZ06}. Interestingly, \totp{}
is the class of self-reducible problems in \sP{} that have a decision version in \cP~\cite{PZ06}; note that prominent \sP-complete problems belong to \totp, such as \ \perfmatch\ and \dnfsat.
Complete problems under parsimonious reductions for \totp{} were provided in~\cite{ABCPPZ22}, e.g.\ \textsc{Size-Of-Subtree}~\cite{K74}.
The significance of \totp{} and its relationship with the class of approximable counting problems have been investigated in~\cite{PZ06,Arenas,BCP20,ABCPPZ22,AC23}. 
\\
The two classes \sP\ and \totp\ imply two paradigms of counting computation models that exhibit significant similarities and differences: on one hand, the computational hardness of computing the exact function value is similar for both models, as shown by the fact that they share complete problems under polynomial-time Turing reductions~\cite{PZ06}; on the other hand, checking whether a solution exists is in \cP\ for all problems in \totp, while it is even \NP-complete for some problems in \sP, a fact that shows that \totp\ is strictly included in \sP, unless $\cP=\NP$. 
\\
In this paper, we build upon the comparison between these two paradigms by studying well-known classes of decision problems, currently defined by means of `accepting-path counting', under the perspective of the `total counting' model. In particular, we consider the complexity classes shown in Table~\ref{classes definitions}, which are defined using conditions on the number of accepting paths, or the difference---which is called the \textit{gap}---between accepting and rejecting paths of an NPTM~\cite{FFK94}; 
for all these classes we introduce their `\totp' counterparts, i.e.\  classes defined by an analogous condition on functions that count the \textit{total number of computation paths} of NPTMs. We compare each `traditionally defined' class with its counterpart, showing that many of them remain the same under both models. We thus obtain alternative characterizations for these classes that lead to novel insights and results on their computational complexity. Notably, we 
provide new complete problems for \op, \modkp, \cep, \pp, \spp\ and \wpp, by using the `total counting' paradigm.\smallskip \\
\begin{table}[t]
\begin{center}
\begin{tabular}{ | m{1.5cm} | m{2.3cm} | m{4.0cm} | m{3.9cm}| } 
\hline
Class & Function $f$ in: & If $x\in L$: & If $x\notin L$:  \\ 
\hline
\up & \sP\ & $f(x)=1$& $f(x)=0$\\
\hline
\fewp & \sP\ & $f(x) \leq p(|x|)$ for some polynomial $p$ and $f(x)>0$ & $f(x)=0$\\
\hline
\op & \sP\ & $f(x)$ is odd & $f(x)$ is even\\
 \hline
\modkp & \sP & $f(x)\not\equiv 0\pmod{k} $ & $f(x)\equiv 0 \pmod{k}$\\
 \hline
 \spp  & \gapp\ & $f(x)=1$ & $f(x) = 0$\\ 
\hline
\wpp & \gapp\ & $f(x)=g(x)$ for some  $g\in\FP$ with $0\not\in~\mathrm{range}(g)$ & $f(x) = 0$\\ 
\hline
\cep & \gapp\ & $f(x)=0$ & $f(x)\neq 0$ [\small{alt-def:} $f(x) > 0$] \\ 
\hline
\pp  & \gapp\ & $f(x)> 0$ & $f(x)\leq 0$ [\small{alt-def:} $f(x) < 0$]\\ 
 \hline
\end{tabular}
\end{center}
\caption{Classes \up~\cite{Valiant76}, \fewp~\cite{AR88}, \op~\cite{PZ83}, \modkp~\cite{CH89,BG92,H90},  \spp~\cite{OH93,FFK94}, \wpp~\cite{FFK94}, \cep~\cite{Simon75}, and \pp~\cite{Simon75,gill}. }
\label{classes definitions}
\end{table}
\noindent\textbf{Related work.}
Interestingly, several of the classes demonstrated in Table~\ref{classes definitions} have attracted attention, as either they have been essential for proving important theorems, or they contain significant problems. 
Specifically, the classes \op\ and \pp\ have received much attention due to their relation to Toda's theorem. 
 Problems in \op\ and \pp\ can be decided with the information of the rightmost and leftmost bit of a \sP\ function, respectively. Toda's theorem consists of two important results. First, \PH\ can be reduced to \op\ under probabilistic reductions, i.e. $\PH\subseteq\textsf{BPP}^\op$. Second, $\textsf{BPP}^\op$ is contained in $\cP^{\pp}$, which in turn implies that is also contained in $\cP^{\sP}$, where one oracle call suffices. In fact, the oracle needs to compute the value of a \sP\ function modulo $2^m$, for some $m$.
 Another prominent result, preceding Toda's theorem, is the Valiant--Vazirani theorem~\cite{VV86}, stating that $\NP\subseteq \RP^{\up}$, which implies that \textsc{Sat} remains hard even if the input instances are promised to have at most one satisfying assignment. 
 \\
 The complexity class \up{} has also been of great significance in cryptography, where the following statement holds: $\cP=\up$ if and only if there are no one-way functions.
 The class \spp\ attracted attention when \textsc{Graph Isomorphism} was shown to lie in it~\cite{AK06}. The  \textsc{Graph Isomorphism} problem is believed to be one of the few  \NP-intermediate problems; there is no known polynomial-time algorithm for it, and there is strong evidence that it is not \NP-complete~\cite{Schoning88}. \spp\ can be seen as the gap-analog of \up. In~\cite{FFK94} it was shown that \spp\ is the smallest reasonable gap-definable class. 
 \\
 Valiant in~\cite{Valiant06} and Curticapean in~\cite{C15} have provided complete problems for \op\ and \cep, respectively, which are defined by counting problems that are not \sP-complete under parsimonious reductions (unless $\cP=\NP$). 
 In particular, Curticapean proved in~\cite{C15} that the problem of determining whether two given graphs have the same number of perfect matchings is complete for \cep. He also proved that this problem has no subexponential algorithm under the Exponential Time Hypothesis (ETH).
 The problem of counting perfect matchings in a graph, namely \perfmatch, is \sP-complete and \totp-complete under poly-time Turing reductions~\cite{Valiant79,PZ06}. However, it is not known to be complete for either of these classes under parsimonious reductions. \smallskip \\
%
\noindent\textbf{Our contribution.} In Section~\ref{gap tot p subsection}, we introduce the classes that are demonstrated in Table~\ref{totp classes} which are defined via \totp\ functions.
As \totp\ is a \textit{proper} subclass of \sP\ (unless \cP=\NP), 
the first interesting question we answer is whether these classes are proper subclasses 
of the corresponding ones shown in Table~\ref{classes definitions}. 
Our results exhibit a dichotomy; these classes are either equal to \cP, namely $\utp=\fewtp=\cP$, or equal to their analogs definable by \sP\ functions (Propositions~\ref{up}, \ref{fewt}, \ref{odd totp = odd p}, \ref{modkp}, and Corollary~\ref{gap tot classes equal to gap classes}). 
\\
The results of Section~\ref{gap tot p subsection} provide an alternative model of computation for \op, \modkp, \spp, \wpp, \cep, and \pp. These results also mean that the `\totp' model captures the essence of the aforementioned classes, while the `\sP' model turns out to be 
somewhat harder than necessary to define them. As a consequence, in Section~\ref{complete problems}, for each of these classes, we obtain a new family of complete problems 
that are defined by \totp-complete problems under parsimonious reductions, which are not \sP-complete under the same kind of reductions unless $\cP=\NP$. Thus, we generalize the completeness results by Valiant and Curticapean for \op\ and \cep, respectively.  In fact, an analogous model of computation and analogous complete problems are obtained for every gap-definable class, and not only the ones mentioned in this work. 
\\
 We also present and study problems defined via the difference of the value of total counting problems.
 Building upon a relevant result by Curticapean~\cite{C15}, we show that such difference problems defined via the \totp{} function \perfmatch{} are complete for the classes \pp{} and \wpp, respectively (Propositions~\ref{pp} and~\ref{wpp}); we also show a hardness result for \spp{} (Proposition~\ref{spp}).
 \\
Finally, in Subsection~\ref{wpp complete}, we prove that under the randomized Exponential Time Hypothesis  (rETH), there is no subexponential algorithm for the promise problem $\textsc{DiffPerfMatch}_{=g}$, which is the problem of determining whether the difference between the number of perfect matchings in two graphs is zero or equal to a specific value.


\section{Preliminaries}\label{preliminaries}


\begin{definition}[\cite{Valiant79,PZ06,FFK94}]\label{classes}
(a) 
$\sP=\{acc_M:\Sigma^*\rightarrow \mathbb{N} \ | \  M\text{ is an NPTM}\}$,

(b) $\FP= \{f:\Sigma^*\rightarrow \mathbb{N} ~\mid~  f\text{ is computable in polynomial time}\}$,
 
	
(c) $\totp =\{tot_M:\Sigma^*\rightarrow \mathbb{N} \ | \  M\text{ is an NPTM}\}$,

(d)  $\gapp=\{\Delta M:\Sigma^*\rightarrow \mathbb{N} \ | \  M\text{ is an NPTM}\}$,\\
where $acc_M(x)=\#($accepting paths of $M$ on input $x)$, $tot_M(x)= \#($all computation paths of $M$ on input $x)-1$, $\Delta M(x)=acc_M(x)-rej_M(x)$, and $rej_M(x)=\#($rejecting paths of $M$ on input $x)$, for every $x\in\Sigma^*$.
\end{definition} 

\begin{remark}
Since every NPTM has at least one computation path, one is subtracted by the total number
of paths in the definition of \totp, so that functions in the class can take the zero value. 
As a result, many natural counting problems lie in \totp. 
\end{remark}

Below, we include a remark regarding some of the classes presented in Table~\ref{classes definitions}.

\begin{remark} [Table~\ref{classes definitions}]
\begin{enumerate}[label=\textit{(\alph*})]
    \item Note that $\ensuremath{\mathsf{Mod_2P} = \op}$ and \modkp\ for $\mathsf{k}>2$ is a generalization of \op, based on congruence mod integers other than two.
    \item The class \modkp\ was defined in~\cite{BG92} via the acceptance condition `$x\in L$ iff $f(x)\not\equiv 0\pmod{k}$'. On the other hand, in~\cite{CH89} \modkp\ was defined via the alternative condition `$x\in L$ iff $f(x)\equiv 0\pmod{k}$' (under which the class of~\cite{BG92} would be $\ensuremath{\mathsf{coMod_k P}}$).
    \item For the alternative definition of \cep, note that given a function $f\in \gapp$, that satisfies the first definition, we have that the function $f^2$ belongs to \gapp\ as well, since \gapp\ is closed under multiplication and $f^2$ satisfies the alternative definition, i.e.\ $x\in L$, if $f^2(x)=0$, whereas $x\not\in L$, if $f^2(x)>0$.
\end{enumerate}
\end{remark}

Various kinds of reductions are used between counting problems. In particular, parsimonious reductions preserve the exact value of the two involved functions.

\begin{definition}  $f$ reduces to  $g$ under \textit{parsimonious reductions}, denoted $f \leq_{\mathrm{par}}^p g$, if and only if there is $h \in \FP$, such that for all  $x\in \Sigma^*$, $f(x)=g(h(x))$.
\end{definition}

Informally, a function is self-reducible if its value on an instance can be recursively computed by evaluating the same function on a polynomial number of smaller instances. The formal definition of (poly-time) self-reducible functions is the following.

\begin{definition}[\cite{ABCPPZ22}]
A function $f : \Sigma^*\rightarrow \mathbb{N}$ is called (poly-time) self-reducible if there exist polynomials 
$r$ and $q$, and polynomial-time computable functions $h : \Sigma^*\times\mathbb{N} \rightarrow \Sigma^*$, 
$g :\Sigma^*\times\mathbb{N} \rightarrow \mathbb{N}$, and $t :\Sigma^*\rightarrow\mathbb{N}$ such that for all 
$x\in\Sigma ^*$:
\begin{enumerate}[label=\textit{(\alph*})]
\item $f$ can be processed recursively by reducing $x$ to a polynomial number of instances $h(x,i)$ $(0 \le i\le r(|x|))$, i.e.\ formally for every $x\in\Sigma^*$  $$f(x) = t(x) +\sum_{i=0}^{r(|x|)} g(x,i)f(h(x,i))$$
\item  the recursion terminates after at most polynomial depth, i.e.\ formally the depth of the recursion is $q(|x|)$, and for every $x\in\Sigma^*$ and $j_1,j_2,\dots,j_{q(|x|)}\in \{0,\dots ,r(|x|)\}$, $f\big(h(\dots h(h(x,j_1),j_2)\dots ,j_{q(|x|)})\big)$ can be computed in polynomial time.
\item every instance invoked in the recursion is of $poly(|x|)$ size, i.e.\ formally for every $x\in\Sigma^*$, $k\leq q(|x|)$ and $j_1,j_2,\dots ,j_k \in \{0,\dots ,r(|x|)\}$ it holds that $$|h(\dots h(h(x,j_1),j_2)\dots ,j_k\big)|\in\mathcal{O}\big(poly(|x|)\big).$$
\end{enumerate}
\end{definition}

\begin{figure}
\centering
\includegraphics[scale=0.30]{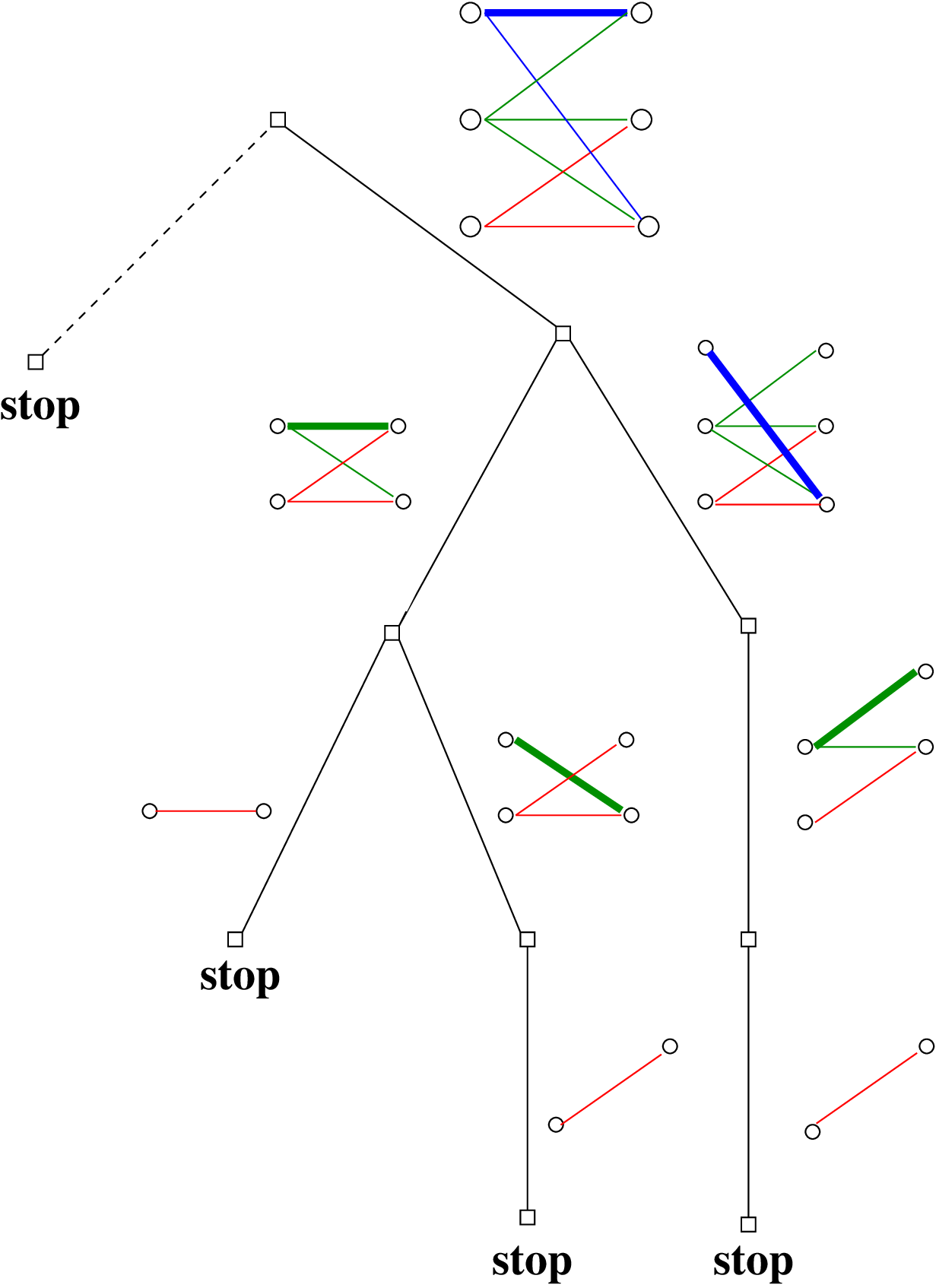}
\vspace{-1pt}
\caption{NPTM $M$ for which it holds that $tot_M(G) = \#(\text{perfect matchings on input } G)$, where $G$ is a bipartite graph~\cite{PZ06}. Thick lines indicate graph edges, based on which a non-deterministic branching is performed, depending on whether there exist perfect matchings that contain the edge (left subtree) and perfect matchings that do not contain it (right subtree).}
\vspace{5pt}
\label{fig:totp}
\end{figure}

Figure~\ref{fig:totp} depicts how self-reducibility and the easy-decision property of counting perfect matchings in a bipartite graph imply that the problem belongs to \totp.

\totp{} is also a \textit{robust} class~\cite{Arenas}; it has natural complete problems~\cite{ABCPPZ22} and nice closure properties as stated in the following proposition. Note that \sP{} is not closed under subtraction by one unless $\spp\subseteq\NP$~\cite{OH93}.

\begin{proposition}\label{closure}
\totp\ is closed under addition, multiplication, and subtraction by one.
\end{proposition}
%
\begin{proof} 
We  show that if $f,g\in\totp$, then $h_1=f+g$, $h_2= f\cdot g$ and $h_3 = f\dot- 1$ also belong to \totp.
Specifically, $h_3:\Sigma^*\rightarrow \mathbb{N}$ is defined by $$h_3(x)=\begin{cases}
f(x)-1, &\text{ if }f(x)\neq 0 \\
f(x), &\text{ if } f(x)=0
\end{cases}.$$

Let $M_f$, $M_g$ be NPTM such that for every $x\in\Sigma^*$, $f(x)=tot_{M_f}(x)=\#($paths of  $M_f$ on  $x)-1$ and $g(x)=tot_{M_g}(x)=\#(\text{paths of } M_g \text{ on } x)-1$. We  construct $M_1$, $M_2$ and $M_3$ such that $h_i(x)=tot_{M_i}(x)=\#(\text{paths of } M_i \text{ on } x)-1$, for $i\in\{1,2,3\}$.
\begin{itemize}
    \item \textit{Subtraction by one}: $M_3$ on $x$ simulates $M_f$ on $x$ with a few modifications as follows. If $M_f$ has only one path, then $M_3$ does exactly what $M_f$ does. If $M_f$ makes at least one non-deterministic choice, $M_3$ copies the behavior of $M_f$, but while simulating the leftmost path, before making a non-deterministic choice, it checks whether one of the choices leads to a deterministic computation. The first time $M_3$ detects such a choice, it eliminates the path corresponding to the deterministic computation and continues the simulation of $M_f$. Notice that $M_3$ can recognize the leftmost path since computation paths can be lexicographically ordered. In this case, $M_3$ has one path less than $M_f$. In both cases, it holds that $h_3(x) = tot_{M_3}(x) = tot_{M_f}(x)\dot-1 = f(x)\dot- 1$.
    \item \textit{Addition}: If one of the machines, let's say $M_f$, has one computation path on input $x$, then $f(x)=0$. So, on input $x$, $M_1$ checks whether either $M_f$ or $M_g$ has exactly one path, and if this the case, it simulates the other one, i.e. $M_g$ or $M_f$, respectively. Otherwise, on input $x$, $M_1$ simulates $M_3$ and $M_g$ non-deterministically, i.e. in two different branches. Since $\#(\text{paths of } M_3\text{ on } x)= h_3(x)+1 = f(x)$ and $\#(\text{paths of } M_g\text{ on } x)= g(x)+1$, we have that $\#(\text{paths of } M_1\text{ on } x)= \#(\text{paths of } M_3\text{ on } x)+\#(\text{paths of } M_g$ \\ $\text{ on } x)= f(x)+g(x)+1$. This implies that $tot_{M_1}(x) = f(x)+g(x)=h_1(x)$.
    \item \textit{Multiplication}: If one of the machines, let's say $M_f$, has one computation path on input $x$, then $f(x)=0$. So, on $x$, $M_2$ checks whether at least one of $M_f$ and $M_g$ has exactly one path, and if this is true, it generates one path and halts. Otherwise, consider the function $h_4:\Sigma^*\rightarrow\mathbb{N}$ such that $h_4(x)=g(x)\dot- 1$ for every $x\in\Sigma^*$, and the NPTM $M_4$ such that $h_4(x)=tot_{M_4}(x)$. On input $x$, $M_2$ generates two branches. The first branch is a dummy path. On the second branch, $M_2$ simulates $M_3$ and $M_4$ sequentially.  So, $\#(\text{paths of } M_2\text{ on } x)=\#(\text{paths of } M_3\text{ on } x)\cdot\#(\text{paths of } M_4\text{ on } x) + 1 =f(x)\cdot g(x)+1=h_2(x)$.
\end{itemize}
\end{proof}


Alternative characterizations of \totp{} and \gapp{} are provided in the following proposition, together with their relation to classes \FP, \pp{} and \sP.

\begin{proposition}[\cite{PZ06,FFK94}]
\label{PZmt}
(a) $\FP \subseteq \totp \subseteq \sP$. The inclusions are proper unless $\cP=\NP$.

(b) \totp{} is the closure under parsimonious reductions of the class of self-reducible \sP{} functions, whose decision version is in \cP, where the decision version of a function $f:\Sigma^*\rightarrow \mathbb{N}$ is $L_f=\{x\in \Sigma^* \mid f(x)>0\}$. 

(c) $\gapp=\sP - \sP = \sP-\FP=\FP-\sP$, where the subtraction of a function class from another, denotes the class of functions that can be described as the difference of two functions, one from each class.

(d) $\cP^{\totp}=\cP^{\sP}=\cP^{\pp}=\cP^\gapp$.
\end{proposition}

The rationale for introducing the class \gapp{} is to capture variants of \sP{} problems that take negative values as well. For example, the permanent of a matrix with non-negative integer entries is a \sP{} problem~\cite{Valiant79}, while the permanent of a matrix with arbitrary, possibly negative entries, lies in \gapp{} \cite{FFK94}.



Next, we provide known relationships among the classes of Table~\ref{classes definitions} in Proposition~\ref{relationships} and the Valiant--Vazirani and Toda's theorems in Theorem~\ref{VV-and-Toda}.

\begin{proposition}[\cite{RRW94}]\label{relationships}
(a) $\up\subseteq \fewp \subseteq \NP \subseteq \mathsf{coC_=P}\subseteq \pp$.

(b) $\fewp\subseteq \spp \subseteq \wpp \subseteq \cep\subseteq \pp$.

(c) $\spp\subseteq \op\subseteq \modkp$.
\end{proposition}


\begin{theorem}[\cite{VV86,Toda91}]\label{VV-and-Toda}
(a) Valiant--Vazirani theorem: $\NP\subseteq \RP^{\up}$.

(b) Toda's theorem: $\PH\subseteq\BPP^{\op}\subseteq \cP^{\sP[1]}$.
\end{theorem}

In Subsection~\ref{wpp complete} we use the Exponential Time Hypothesis and its randomized variant, namely rETH, which are given below.

\begin{itemize}
    \item ETH: There is no deterministic algorithm that can decide 3-\SAT\ in time $\exp(o(n)).$
    \item rETH: There is no randomized algorithm that can decide 3-\SAT\ in time $\exp(o(n)),$ with error probability at most $1/3.$
\end{itemize}

\section{Classes defined by total counting}\label{gap tot p subsection}

\subsection{The class \gaptp}

\begin{definition}\label{def:gaptp}
A function $f$ belongs to the class \gaptp\ iff it is the difference of two \totp\ functions.
\end{definition}

Proposition~\ref{gap tot p} demonstrates that \gaptp\ coincides with the class \gapp{}. Corollary~\ref{gaptp} provides alternative characterizations of \gapp\ and \gaptp.

\begin{proposition}\label{gap tot p}
$\gaptp=\gapp$.
\end{proposition}
\begin{proof}
$\gaptp\subseteq \gapp$ is straightforward, since $\totp\subseteq\sP$. For $\gapp\subseteq \gaptp$, note that for any \sP\ function $f$, there exist NPTMs $M$, $M'$ such that $f(x)=acc_M(x)=tot_{M'}(x)-tot_M(x)$, where we obtain $M'$ by doubling the accepting paths of $M$. So for any $g\in\gapp$ there exist NPTMs $N$, $N'$, $M$, $M'$ such that $g(x)=acc_N(x)-acc_M(x) = (tot_{N'}(x)-tot_N(x))-(tot_{M'}(x)-tot_M(x))= (tot_{N'}(x)+tot_M(x))-(tot_N(x)+tot_{M'}(x))=tot_{M_1}(x)-tot_{M_2}(x)$, where $M_1$, $M_2$ can be constructed as described in the proof of Proposition~\ref{closure}.
\end{proof}

\begin{corollary}\label{gaptp}
 $\gaptp = \gapp = \sP - \sP = \totp - \totp = \sP - \FP = \FP - \sP = \FP - \totp = \totp - \FP$.
 \end{corollary}
 \begin{proof} 
The first five equalities follow from Propositions~\ref{PZmt}(c), \ref{gap tot p}, and Definition~\ref{def:gaptp}.  To prove $\FP - \sP = \FP - \totp$, we show that $\FP - \sP \subseteq \FP - \totp$.  The inverse inclusion $\FP - \totp \subseteq\FP - \sP$ is trivial, since $\totp\subseteq\sP$. Analogously, we can show that  $\sP - \FP \subseteq \totp-\FP$.

 Let $g\in\FP$ and $f=acc_M \in\sP$. Then, $g(x)-f(x) = g(x) - (tot_{M'}(x) - tot_M(x))$, where $M'$ is obtained from $M$ by doubling its accepting paths (as in the proof of Proposition~\ref{gap tot p}). W.l.o.g.\ we assume that the computation tree of $M$ is a perfect binary tree (or in other words, $M$ is in normal form), so we have that $g(x) - (tot_{M'}(x) - tot_M(x)) = g(x) - (tot_{M'}(x) - (2^{p(|x|)} -1)) = g'(x) - tot_{M'}(x)$, where  $g'\in\FP$ with $g'(x) = g(x) + 2^{p(|x|)}-1$ for every $x$, where $p(|x|)$ is the polynomial bound on the length of $M(x)$'s paths.
\end{proof}

\smallskip
 The above corollary demonstrates, among other implications, that \gaptp{} contains problems in $\FP - \totp$, such as counting the unsatisfying assignments of a formula in DNF, which is not in \totp{} unless $\cP=\NP$.


\subsection{The classes \utp, \fewtp, \otp, \modktp, \sppt, \wppt, \cetp, and \ppt}\label{tot definable classes}

\begin{table}[ht]
\begin{center}
\begin{tabular}{ | m{1.5cm} | m{2.3cm} | m{4.0cm} | m{3.9cm}| } 
\hline
Class & Function $f$ in: & If $x\in L$: & If $x\notin L$:  \\ 
\hline
\utp\ & \totp\ & $f(x)=1$ & $f(x)=0$\\
\hline
\fewtp\ & \totp\ & $f(x) \leq p(|x|)$ for some polynomial $p$ and $f(x)>0$ & $f(x)=0$\\
\hline
\otp\ & \totp\ & $f(x)$ is odd & $f(x)$ is even\\
 \hline
 \modktp\ & \totp\ & $f(x)\not\equiv 0\pmod{k} $ & $f(x)\equiv 0\pmod{k}$\\
 \hline
 \sppt\ & \gaptp\ & $f(x)=1$ & $f(x) = 0$\\ 
 \hline
 \wppt\ & \gaptp\ & $f(x)=g(x)$ for some $g\in\FP$ with\quad $0\not\in~\mathrm{range}(g)$ & $f(x) = 0$\\
 \hline
\cetp\ & \gaptp\ & $f(x)=0$ & $f(x)\neq 0$ [\small{alt-def:} $f(x) > 0$]\\ 
 \hline
 \ppt\  & \gaptp\ & $f(x)> 0$ & $f(x)\leq 0$ [\small{alt-def:} $f(x) < 0$]\\ 
\hline
\end{tabular}
\end{center}
\caption{Classes \utp, \fewtp, \otp, \modktp, \sppt, \wppt, \cetp, \ppt.}
\label{totp classes}
\end{table}
The classes defined in Table~\ref{totp classes} are the \totp-analogs of those contained in Table~\ref{classes definitions}.
First, we show that $\cP=\utp=\fewtp$ and next, that every other class of Table~\ref{totp classes} coincides with its counterpart from Table~\ref{classes definitions}.

\begin{proposition}\label{up}
(a) $\cP = \utp\subseteq \up$.

(b)  If $\up\subseteq \utp$, then $\cP = \up$ (and thus $\RP=\NP$).
\end{proposition}
\begin{proof}
(a) $\utp\subseteq\cP$: Let $L\in\utp$. Then there exists an NPTM $M$ such that $x\in L$ iff $M$ has 2 paths, whereas $x\not\in L$ iff $M$ has 1 path. Define the polynomial-time Turing machine $M'$ that on any input, simulates either the unique path or the two paths of $M$ deterministically, and it either rejects or accepts, respectively. To prove the inverse inclusion $\cP\subseteq\utp$, consider a language $L\in \cP$ and define the NPTM $N$, which, on any input $x$, simulates the deterministic polynomial-time computation for deciding $L$ and generates one or two paths if the answer is negative or positive, respectively. The inclusion $\utp\subseteq \up$ is immediate from $\totp\subseteq\sP$.
(b) From (a), if $\up\subseteq \utp$, then $\up\subseteq \cP$ and $\RP=\NP$ by the Valiant--Vazirani theorem (Theorem~\ref{VV-and-Toda}(a)).
\end{proof}

\begin{proposition}\label{fewt}
(a) $\cP=\utp=\fewtp$.\\
(b) If $\fewp\subseteq \fewtp$, then $\cP = \fewp$.
\end{proposition}
\begin{proof} 
Part (b) is immediate from (a). For (a), let $L\in\fewtp$. Consider the Turing machine $M$ which has either more than 1 but polynomially many paths if $x\in L$  or just 1 path if $x\not\in L$. Define $M'$ to be the NPTM that on any $x$ deterministically simulates all the paths of $M$ on $x$ and in the first case, $M'$ makes a branching forming two paths, while in the second case, it halts. So, $L\in\utp$.
\end{proof}

\smallskip

In fact, a stronger fact than Propositions~\ref{up} and~\ref{fewt} is true. 

\begin{proposition}\label{totp-poly-paths}
Let $f\in\totp$. If $f(x)\leq p(|x|)$, for every $x\in\Sigma^*$, where $p$ is a polynomial, then $f\in\FP$. 
\end{proposition}
%
%
\begin{proof} 
The proof follows that of Propositions~\ref{up} and~\ref{fewt}. Let $f\in\totp$ and $f(x)\leq p(|x|)$, for every $x\in\Sigma^*$ and some polynomial $p$; let also $M$ be the NPTM such that $f(x)=\#(\text{paths of } M \text{ on input } x)-1$. Consider the deterministic polynomial-time Turing machine $M'$ that on input $x$ sets a counter to zero, simulates all the paths of $M$, and increases the counter by one each time a path comes to an end. Finally, $M'$ outputs the final result of the counter minus one.
\end{proof}

\smallskip

The class \otp\ (odd-P or parity-P) is the class of decision problems, for which the acceptance condition is that the number of all computation paths of an NPTM is even (or the number of all computation paths minus 1 is odd).


Valiant provided in~\cite{Valiant06} an \op-complete problem definable by a \totp{} function. 
Let $\oplus\textsc{Pl-Rtw-Mon-3CNF}$ be the problem that on input a planar 3CNF formula where each variable appears positively and in exactly two clauses, accepts iff the formula has an odd number of satisfying assignments. The counting version of this problem, namely \#\textsc{Pl-Rtw-Mon-3CNF}, is in \totp; it is self-reducible like every satisfiability problem and has a decision version in \cP, since every monotone formula has at least one satisfying assignment.

\begin{proposition}[\cite{Valiant06}] \label{odd planar mon}
$\oplus\textsc{Pl-Rtw-Mon-3CNF}$ is \op-complete.
\end{proposition}

\begin{proposition}\label{odd totp = odd p}
$\otp=\op$.
\end{proposition}
\begin{proof}
$\otp\subseteq\op$: Consider a language $L\in\otp$ and the NPTM $M$ such that $x\in L$ iff $tot_M(x)$ is odd. Consider an NPTM $M'$ that on any input $x$, simulates $M$ on $x$. Since, $M'$ can distinguish the leftmost path of $M$, it rejects on this path, and it accepts on every other path.  Then, $x\in L$ iff $acc_{M'}(x)$ is odd.

$\op\subseteq\otp$: Let $L\in\op$ and $M$ be the NPTM such that $x\in L$ iff $acc_M(x)$ is odd. Then, we obtain $M'$ by doubling the rejecting paths of $M$ and adding one more path. It holds that $x\in L$ iff $tot_{M'}(x)$ is odd.
\end{proof}

\medskip
\begin{altproof} 
Let $L\in\op$. Then, it holds that $x\in L$ iff $f(x)\equiv 1 \pmod{2}$ for some $f\in\sP$. By Proposition~\ref{odd planar mon}, there is an $h\in\FP$, such that $x\in L$ iff $\#\textsc{Pl-Rtw-Mon-3CNF}(h(x)) \equiv 1 \pmod{2}$. So, define the \totp\ function $g = \#\textsc{Pl-Rtw-Mon-3CNF}\circ h$. Then $x\in L$ iff $g(x)\equiv 1 \pmod{2}$ and thus $L\in\otp$.
\end{altproof}

\begin{proposition}\label{modkp}
$\modktp=\modkp$.
\end{proposition}
\begin{proof} 
The proof of $\modktp\subseteq\modkp$ is very similar to the proof of $\otp\subseteq\op$ in Proposition~\ref{odd totp = odd p}. For the inclusion $\modkp\subseteq \modktp$, let $L\in\modkp$ and $M$ be the NPTM such that $x\in L$ iff $acc_M(x)\equiv a \pmod{k}$, for some $a\in\{1,...,k-1\}$. Then, $M'$ can be obtained from $M$ by generating $k$ paths for every rejecting path and one more path (a dummy path). So, $\#(\text{paths of } M'\text{ on input } x)-1 \equiv a \pmod{k}$. So, there is an NPTM $M'$, such that $tot_{M'}(x) \equiv acc_M(x) \pmod{k}$.
\end{proof}

\begin{remark}
The above proof shows that not only equivalence or non-equivalence modulo $k$, but also the value of the \sP\ function modulo $k$ is preserved.
\end{remark} 

\begin{remark}
So, we can say that if we have information about the rightmost bit of a \totp\ function is as powerful as having information about the rightmost bit of a \sP\ function. Toda's theorem would be true if we used \otp\ instead of \op. Moreover, it holds that $\ensuremath\mathsf{BPP}^\op\subseteq \cP^{\totp[1]}$, where it suffices to make an oracle call to a \totp\ function mod $2^m$, for some $m$. However, \utp\ is defined by a constraint on a \totp\ function that yields only NPTMs with polynomially many paths. This means that \utp\  gives no more information than the class \cP\ and as a result, it cannot replace the class \up\ in the Valiant--Vazirani theorem.
\end{remark}

By Proposition~\ref{gap tot p} and the definitions of the classes \sppt, \wppt, \cetp, and \ppt, we obtain the following corollary. 

\begin{corollary}\label{gap tot classes equal to gap classes}
(a) $\sppt = \spp$, (b) $\wppt = \wpp$, (c) $\cetp=\cep$, and (d) $\ppt=\pp$.
\end{corollary}

A more general corollary of Proposition~\ref{gap tot p} is that every gap-definable class coincides with its \totp-analog.

The next corollary is an analog of Proposition~\ref{relationships}. It is an immediate implication from Propositions~\ref{relationships}, \ref{up}, \ref{fewt}, \ref{odd totp = odd p}, \ref{modkp}, and Corollary~\ref{gap tot classes equal to gap classes}.

\begin{corollary}
(a) $\cP = \utp = \fewtp \subseteq \NP \subseteq \mathsf{co}\cetp \subseteq \ppt$.

(b) $\fewtp\subseteq \sppt \subseteq \wppt \subseteq \cetp\subseteq \ppt$.

(c) $\sppt\subseteq \otp\subseteq \modktp$.
\end{corollary}


\section{Hardness results for problems definable via \totp\ functions}\label{complete problems}

In this section, we introduce a new family of complete problems for \op, \modkp, and gap-definable classes. 

\begin{definition}\label{problems-def}
Given a \sP\ function $\#\textsc{A}:\Sigma^*\rightarrow\mathbb{N}$, we define the following decision problems associated with $\#\textsc{A}$:
\begin{itemize}
    \item $\oplus\textsc{A}$ which on input $x\in\Sigma^*$, decides whether $\#\textsc{A}(x)$ is odd.
    \item $\textsc{Mod}_k\textsc{A}$ which on input $x\in\Sigma^*$, decides whether $\#\textsc{A}(x) \not\equiv 0 \pmod{k}$.
    \item $\textsc{DiffA}_{=0}$ which on input $(x,y)\in\Sigma^*\times\Sigma^*$, decides whether $\#\textsc{A}(x)=\#\textsc{A}(y)$.
    \item $\textsc{DiffA}_{>0}$ which on input $(x,y)\in\Sigma^*\times\Sigma^*$, decides whether $\#\textsc{A}(x)>\#\textsc{A}(y)$.
    \item the promise problem $\textsc{DiffA}_{=1}$ which on input $(x,y)\in I_{YES}\cup I_{NO}$, decides whether $(x,y)\in I_{YES}$, where 
    $I_{YES}=\{(x,y) \mid \#\textsc{A}(x) = \#\textsc{A}(y) + 1\}$ and
     $I_{NO}=\{(x,y) \mid \#\textsc{A}(x) = \#\textsc{A}(y)\}$, $x,y\in\Sigma^*$.
     \item the promise problem $\textsc{DiffA}_{=g}$ which on input $(x,y,k)\in I_{YES}\cup I_{NO}$, decides whether $(x,y,k)\in I_{YES}$, where $I_{YES}=\{(x,y,k) \mid \#\textsc{A}(x) = \#\textsc{A}(y) + k\}$ and
     $I_{NO}=\{(x,y,k) \mid \#\textsc{A}(x) = \#\textsc{A}(y)\}$, $x,y\in\Sigma^*$, $k\in\mathbb{N}_{>0}$.
\end{itemize}
\end{definition}

\begin{proposition}\label{problems}
 For any function $\#\textsc{A}\in\sP$, it holds that:
\begin{enumerate}[label=\textit{(\alph*})]
\item $\oplus\textsc{A}\in\op$, $\textsc{Mod}_k\textsc{A}\in\modkp$, $\textsc{DiffA}_{=0}\in\cep$, $\textsc{DiffA}_{>0}\in\pp$, $\textsc{DiffA}_{=1}\in\spp$, and $\textsc{DiffA}_{=g}\in\wpp$.
 \item If $\#\textsc{A}$ is 
 \sP-complete or \totp-complete under parsimonious reductions, then $\oplus\textsc{A}$, $\textsc{Mod}_k\textsc{A}$, $\textsc{DiffA}_{=0}$, $\textsc{DiffA}_{>0}$, $\textsc{DiffA}_{=1}$, and $\textsc{DiffA}_{=g}$ are complete for \op, \modkp, \cep, \pp, \spp\ and \wpp, respectively.
\end{enumerate} 
\end{proposition}
\begin{proof}
 We prove the proposition for the problems $\textsc{DiffA}_{=0}$ and $\textsc{DiffA}_{=g}$. The proof for the other problems is completely analogous.
\begin{enumerate}[label=\textit{(\alph*})]
\item We have that an instance $(x,y)$ of $\textsc{DiffA}_{=0}$ is a $yes$ instance iff $\#\textsc{A}(x)=\#\textsc{A}(y)$ iff $\#\textsc{A}(x)-\#\textsc{A}(y)=0$. The difference $\#\textsc{A}(x)-\#\textsc{A}(y)$ is a \gapp\ function, since it can be written as $\#\textsc{A}'(x,y) - \#\textsc{A}''(x,y)$, where $\#\textsc{A}'(x,y)$ (resp.\ $\#\textsc{A}''(x,y)$) is the function $\#\textsc{A}$ on input $x$ (resp.\ $y$), which means that $\#\textsc{A}',\#\textsc{A}''\in\sP$. 

Similarly, an instance $(x,y,k)$, $k\in\mathbb{N}_{>0}$, of $\textsc{DiffA}_{=g}$ is a $yes$ instance if $\#\textsc{A}(x) - \#\textsc{A}(y) = k$, or equivalently, if $\#\textsc{A}'(x,y,k) - \#\textsc{A}''(x,y,k)= g(x,y,k)$, where $\#\textsc{A}'(x,y,k)$ (resp.\ $\#\textsc{A}''(x,y,k)$) is the function $\#\textsc{A}$ on input $x$ (resp.\ $y$), and $g$ is such that $g(x,y,k)=k$. On the other hand, an instance $(x,y,k)$, $k\in\mathbb{N}_{>0}$, is a $no$ instance if $\#\textsc{A}'(x,y,k) - \#\textsc{A}''(x,y,k)= 0$. Since $\#\textsc{A}',\#\textsc{A}''\in\sP$ and $g\in\FP$ with $0\not\in\mathrm{range}(g)$, the definition of \wpp{} is satisfied and $\textsc{DiffA}_{=g}\in\wpp$.

\item Let $\#\mathsf{A}$ be \totp-complete under parsimonious reductions. By Corollary~\ref{gap tot classes equal to gap classes}(c), a language $L\in\cep$ can be decided by the value of a function $f\in\gaptp$: $x\in L$ iff $f(x) = 0$. By the definition of \gaptp, $f(x)=g(x)-h(x)$ for some $g,h\in\totp$. By \totp-completeness of $\#\mathsf{A}$, we have that $g(x)=\#\textsc{A}(t_1(x))$ and $h(x)=\#\textsc{A}(t_2(x))$, for some $t_1,t_2\in\FP$. So, $x\in L$ iff $g(x)-h(x)=0$ iff $\#\textsc{A}(t_1(x))-\#\textsc{A}(t_2(x))=0$ iff $(t_1(x),t_2(x))$ is a $yes$ instance for $\textsc{DiffA}_{=0}$. 

Likewise, if $L'\in\wpp$, by Corollary~\ref{gap tot classes equal to gap classes}(b), there are $h_1,h_2\in\totp$, such that $x\in L$ if $h_1(x)-h_2(x)=f(x)$, for some $f\in\FP$ with $0\not\in\mathrm{range}(f)$. Then, $x\in L$ if $\#A(t_1(x))-\#A(t_2(x))=f(x)$, for some $t_1,t_2\in\FP$. Equivalently, $x\in L$ if $(t_1(x),t_2(x),f(x))$ is a $yes$ instance of $\textsc{DiffA}_{=g}$. The same reasoning applies to $no$ instances.

In the case of $\#\mathsf{A}$ being \sP-complete under parsimonious reductions, the proof is analogous.
\end{enumerate}
\end{proof}


\begin{example}
For example, the problem $\textsc{DiffSat}_{=0}$ is complete for the class \cep. Note that this problem was defined in~\cite{C15}, where it is denoted by $\textsc{Sat}_{=}$. We use a slightly different notation here, which we believe is more suitable for defining problems that lie in other gap-definable classes.
The problem $\textsc{DiffSat}_{=g}$ takes as input  
two CNF formulas $\phi,\phi'$
and a non-zero natural number $k$, such that either they have the same number of satisfying assignments or the first one has $k$ more satisfying assignments than the second one. The problem is to decide which is the case. This is a generalization of the problem \textsc{Promise-Exact-Number-Sat} defined in~\cite{PC18}. 
\end{example}

By Proposition~\ref{PZmt}(a) and the closure of \totp\ under parsimonious reductions ($\leq_{\mathrm{par}}^p$), if $\cP\neq \NP$, \totp-complete and \sP-complete problems under $\leq_{\mathrm{par}}^p$ form disjoint classes. 
By combining that fact with Proposition~\ref{problems}(b), we obtain a family of complete problems for the classes \op, \modkp, \cep, \pp, \spp, and \wpp{} defined by functions that are \totp-complete under $\leq_{\mathrm{par}}^p$. As a concrete example, consider the particularly interesting problem \sizeofsubtree{}, first introduced by Knuth~\cite{K74} as the problem of estimating the size of a backtracking tree, which is the tree produced by a backtracking procedure. This problem has been extensively studied from an algorithmic point of view (see e.g.~\cite{Purdom78,Chen92}) and was recently shown to be \totp-complete under $\leq_{\mathrm{par}}^p$~\cite{ABCPPZ22}. Proposistion~\ref{problems} implies that the six problems defined via \sizeofsubtree{} as specified in Definition~\ref{problems-def}, are complete for \op, \modkp, \cep, \pp, \spp\ and \wpp, respectively.

Note that, these results provide the first complete problems for \modkp, \spp, \wpp, and \pp\ that are not definable via \sP-complete (under $\leq_{\mathrm{par}}^p$) functions. Moreover, as every gap-definable class coincides with its \totp-analog, any such class has complete problems defined by \totp-complete problems under $\leq_{\mathrm{par}}^p$. Alternatively, one can say that these complete problems are defined by problems in \cP, and not $\NP$-complete ones (unless $\cP=\NP$).

\subsection{Problems definable via the difference of counting perfect matchings}

Curticapean proved in~\cite{C15} that $\textsc{DiffPerfMatch}_{=0}$ is \cep-complete.  We provide analogous results for the classes \pp{} and \wpp. Note that \perfmatch{} is in \totp, and it is not known to be either \sP-complete or \totp-complete under parsimonious reductions. This is yet another approach to obtain complete problems for \pp{} and \wpp, the counting versions of which are not even known to be \totp-complete.

\begin{proposition} [\cite{C15}]\label{Curt}
 $\textsc{DiffPerfMatch}_{=0}$ is complete for \cep.
\end{proposition}
\begin{proof} 
In~\cite{C15} a reduction from $\textsc{DiffSat}_{=0}$ to $\textsc{DiffPerfMatch}_{=0}$ is described. Given a pair of 3CNF formulas $(\phi,\phi')$, two unweighted graphs $G$, $G'$ can be constructed such that 
$$\sSAT(\phi)-\sSAT(\phi')=2^{-T}(\perfmatch(G)-\perfmatch(G'))$$
\noindent where $T\in\mathbb{N}$ can be computed in polynomial time with respect to the input $(\phi,\phi')$.
\end{proof}

The proofs of Propositions~\ref{pp} and~\ref{wpp} are established by adapting the proof of Proposition~\ref{Curt} given in~\cite{C15}.

\begin{proposition}\label{pp}
$\textsc{DiffPerfMatch}_{>0}$ is complete for \pp.
\end{proposition}
\begin{proof}
The reduction from $\textsc{DiffSat}_{=0}$ to  $\textsc{DiffPerfMatch}_{=0}$ provided in~\cite{C15}, is also a reduction from $\textsc{DiffSat}_{>0}$ to  $\textsc{DiffPerfMatch}_{>0}$.
\end{proof}

\begin{proposition}\label{wpp}
$\textsc{DiffPerfMatch}_{=g}$ is complete for \wpp.
\end{proposition}
\begin{proof} 
By Proposition~\ref{problems}, $\textsc{DiffSat}_{=g}$ is \wpp-complete. We show that $\textsc{DiffSat}_{=g}$ reduces to $\textsc{DiffPerfMatch}_{=g}$. Let $(\phi,\phi',k)$ be an input  to $\textsc{DiffSat}_{=g}$, such that $(\phi,\phi',k)$ is a $yes$ instance if $\sSAT(\phi)-\sSAT(\phi')=k$, where $k\in\mathbb{N}_{>0}$. $\textsc{DiffSat}_{=g}$ reduces to  $\textsc{DiffPerfMatch}_{=g}$ on input $(G,G',l)$, where $G$, $G'$ are the graphs described in the reduction that proves Proposition~\ref{Curt} and $l=2^T\cdot k$, where $T$ is computable in polynomial time. It holds that $(\phi,\phi',k)\in \textsc{DiffSat}_{=g}$ if 
$\sSAT(\phi)-\sSAT(\phi')= k \text{\,\, iff }$ 
$
\perfmatch(G)-\perfmatch(G') = 2^T \cdot (\sSAT(\phi)-\sSAT(\phi'))
= 2^T \cdot k= l.
$
Also, $(\phi,\phi',k)\not\in \textsc{DiffSat}_{=g}$ if
$\sSAT(\phi)-\sSAT(\phi')= 0 \text{\,\, iff }$
$\perfmatch(G)-\perfmatch(G') = 2^T \cdot (\sSAT(\phi)-\sSAT(\phi')) = 0.$
\end{proof}


\smallskip
In contrast, we cannot prove \spp-completeness for
$\textsc{DiffPerfMatch}_{=1}$. However, we can prove that $\textsc{DiffPerfMatch}_{=g}$ is hard for \spp.

\begin{proposition}\label{spp}
The problem $\textsc{DiffSat}_{=1}$ is reducible to $\textsc{DiffPerfMatch}_{=g}$.
\end{proposition}
\begin{proof}
Consider two 3CNF formulas $(\phi,\phi')$, with $n$ variables and $m=\mathcal{O}(n)$ clauses, such that either $\sSAT(\phi)-\sSAT(\phi')=1$ or  $\sSAT(\phi)-\sSAT(\phi')=0$. 

Then, using the polynomial-time reduction of~\cite{C15} two graphs $G$, $G'$ can be constructed such that $$\perfmatch(G)-\perfmatch(G') = c^{|V|} \cdot (\sSAT(\phi)-\sSAT(\phi'))$$ where $|V|=\max\{|V(G)|,|V(G')|\}\}=\mathcal{O}(n+m)$  and $c\in (1,2)$ is a constant depending on $\phi,\phi'$ that can be computed in polynomial time. Moreover, the graphs $G$ and $G'$ have $\mathcal{O}(|V|)$ edges.

So, $\textsc{DiffSat}_{=1}$ on input $(\phi,\phi')$ can be reduced to $\textsc{DiffPerfMatch}_{=g}$ on input $(G,G',c^{|V|})$, where $c\in (1,2)$.
\end{proof}

\vspace{2mm}

According to the proof of Proposition~\ref{spp}, the smallest possible non-zero difference between the number of satisfying assignments of two given 3CNF formulas can be translated to an exponentially large difference between the number of perfect matchings of two graphs. In addition, this exponentially large number depends on the input and it can be efficiently computed. 

Moreover, the aforementioned propositions along with Curticapean's result yield alternative proofs of $\wppt=\wpp$, $\cetp=\cep$, and $\ppt=\pp$.

 \medskip

\noindent\textbf{Corollary~\ref{gap tot classes equal to gap classes} (restated).} \textit{(b) $\wppt = \wpp$, (c) $\cetp=\cep$, and (d) $\ppt=\pp$.}

\medskip
\begin{altproof} 
(c) Let $L\in\cep$. Then, $x\in L$ iff ~ $\sSAT(h_1(x))-\sSAT(h_2(x))=0$ iff $\perfmatch(h_3(x)) - \perfmatch(h_4(x))=0$, for some $h_i\in\FP$, $1\leq i \leq 4$. So, define the \totp\ functions $f_1 = \perfmatch\circ h_3$ and $f_2 = \perfmatch\circ h_4$. Then $f_1$, $f_2$ are \totp\ functions and we have that $x\in L$ iff $f_1(x)-f_2(x)=0$.

The proofs of (b) and (d) are completely analogous.
\end{altproof}

\subsection{An exponential lower bound for the problem $\textsc{DiffPerfMatch}_{=g}$}\label{wpp complete}

Curticapean showed that under ETH, the problem $\textsc{DiffPerfMatch}_{=0}$ has no $2^{o(m)}$ time algorithm on simple graphs with $m$ edges~\cite[Theorem 7.6]{C15}. The proof is based on the fact that the satisfiability of a 3CNF formula $\phi$ is reducible to the difference of \perfmatch\ on two different graphs, such that the number of perfect matchings of the graphs is equal iff $\phi$ is unsatisfiable. The reduction follows the steps of the reductions that are used in the proofs of Propositions~\ref{pp} and~\ref{wpp}.
%
%
%
Using the reduction of Proposition~\ref{spp}, we prove the following corollary. 

\begin{corollary}\label{lower-bound}
Under rETH there is no randomized $\exp(o(m))$ time algorithm for  $\textsc{DiffPerfMatch}_{=g}$ on simple graphs with $m$ edges.
\end{corollary}
\begin{proof}
Given rETH we cannot decide whether a given 3USat formula $\phi$ with $m$ clauses is satisfiable using a randomized algorithm that runs in time $\exp(o(m))$~\cite{CIK+08}. By applying the reduction described in~\cite[Lemma 7.3]{C15} we can construct two unweighted graphs $G$ and $G'$ with $\mathcal{O}(m)$ vertices and edges, such that
\begin{itemize}
    \item if $\phi$ is unsatisfiable, then $\perfmatch(G)-\perfmatch(G') = 0$,
    \item if $\phi$ is satisfiable, then $\perfmatch(G)-\perfmatch(G') = c^{|V|}$, where $|V|=\max\{|V(G)|,|V(G')|\}$ and $c\in (1,2)$ is a constant that depends on the input and can be computed in polynomial time.
\end{itemize}
So, \textsc{3USat} on $\phi$ can be reduced to $\textsc{DiffPerfMatch}_{=g}$ on input $(G,G',c^{|V|})$, for some $c\in (1,2)$. Thus, an $\exp (o(m))$ time randomized algorithm for the problem $\textsc{DiffPerfMatch}_{=g}$ would contradict rETH.
\end{proof}

\begin{remark}\label{spp algo}
A different way to read Proposition~\ref{spp} is the following: a
positive result for \perfmatch\ would imply a corresponding positive result for $\textsc{DiffPerf}$\\ $\textsc{Match}_{=g}$ and therefore, for $\textsc{DiffSat}_{=1}$. Of course, this positive result would be an exponential-time algorithm for these problems!
For example, a fully polynomial approximation scheme for $\perfmatch$ would yield an algorithm that distinguishes between $\perfmatch(G)-\perfmatch(G') = c^{n} \text{ and }$
$\textsc{Perf}$ \\ $\textsc{Match}(G)-\perfmatch(G') =0$ with high probability in time $\mathcal{O}(\frac{2^m}{c^n})$, where  $n=\max\{|V(G)|,|V(G')|\} \text{ and } m=\max\{|E(G)|,|E(G')|\}=\mathcal{O}(n).$ So, in time $\mathcal{O}(d_1^n)$, where $d_1\in(1,2)$.  Note that $d_1$ depends on the input, so this is not a robust result, and that is why we state it just as a remark. 
The same kind of algorithm would then exist for all the problems in \spp. Among them is the well-studied \textsc{Graph Isomorphism}~\cite{AK06}, which is one of the \NP\ problems that have been proven neither \NP-complete, nor polynomial-time solvable so far~\cite{KST93,Babai16}, and all the problems in \up, since $\up\subseteq\spp$ (see Proposition~\ref{relationships}).
\end{remark}

\section{Conclusion}

Our work aims to gain more insights and a better understanding of aspects related to the 
power of the \totp\ model of computation. The contribution of this paper is primarily conceptual but also illustrates how the introduction of appropriate definitions can lead to nontrivial results in a fairly straightforward way, circumventing complex and hard-to-read proofs. The introduction of \totp-analogs of the classes shown in Table~\ref{classes definitions} led to new characterizations and complete problems for \op, \modkp, and all gap-definable classes. The \totp\ computational model was proven to be sufficient, and it is arguably more appropriate to define these classes. Moreover, to the best of our knowledge, for \modkp, \spp, \wpp, and \pp, we present the first complete problems that are not defined via \sP-complete (under $\leq^p_{\mathrm{par}}$) problems.
Finally, two significant results of our approach are (a) that if the randomized Exponential Time Hypothesis holds, the 
\wpp-complete problem
$\textsc{DiffPerfMatch}_{=g}$ has no subexponential algorithm and (b)
 every \spp\ problem (including \textsc{Graph Isomorphism}) is decidable by the difference of \perfmatch\ on two graphs, which is promised to be either exponentially large or zero.
 We expect that our results may inspire further research on total counting functions, as well as on the complexity classes that can be defined via them, thus providing new tools for analyzing the computational complexity of interesting problems that lie in such classes.

\subsubsection{Aknowledgements.} Aggeliki Chalki has been funded by the project “Mode(l)s of Verification and Monitorability” (MoVeMnt) (grant no 217987). Sotiris Kanellopoulos and Aris Pagourtzis have been partially supported for this work by project MIS 5154714 of the National Recovery and Resilience Plan Greece 2.0 funded by the European Union under the NextGeneration EU Program.

\medskip

\bibliographystyle{plain}
\bibliography{bibliography}

\begin{thebibliography}{10}

\bibitem{AC23}
Antonis Achilleos and Aggeliki Chalki.
\newblock Counting computations with formulae: Logical characterisations of counting complexity classes.
\newblock In {\em Proc. of the 48th International Symposium on Mathematical Foundations of Computer Science, {MFCS} 2023}, volume 272 of {\em LIPIcs}, pages 7:1--7:15, 2023.

\bibitem{AR88}
Eric Allender and Roy~S. Rubinstein.
\newblock P-printable sets.
\newblock {\em {SIAM} Journal on Computing}, 17(6):1193--1202, 1988.

\bibitem{ABCPPZ22}
Antonis Antonopoulos, Eleni Bakali, Aggeliki Chalki, Aris Pagourtzis, Petros Pantavos, and Stathis Zachos.
\newblock Completeness, approximability and exponential time results for counting problems with easy decision version.
\newblock {\em Theoretical Computer Science}, 915:55--73, 2022.

\bibitem{Arenas}
Marcelo Arenas, Martin Mu{\~{n}}oz, and Cristian Riveros.
\newblock Descriptive complexity for counting complexity classes.
\newblock {\em Logical Methods in Computer Science}, 16(1), 2020.

\bibitem{AK06}
Vikraman Arvind and Piyush~P. Kurur.
\newblock Graph isomorphism is in {SPP}.
\newblock {\em Information and Computation}, 204(5):835--852, 2006.

\bibitem{Babai16}
L\'{a}szl\'{o} Babai.
\newblock Graph isomorphism in quasipolynomial time [extended abstract].
\newblock In {\em Proc. of the 48th Annual ACM Symposium on Theory of Computing}, STOC~'16, pages 684--697, 2016.

\bibitem{BCP20}
Eleni Bakali, Aggeliki Chalki, and Aris Pagourtzis.
\newblock Characterizations and approximability of hard counting classes below {\#}{P}.
\newblock In {\em Proc. of the 16th International Conference on Theory and Applications of Models of Computation, {TAMC} 2020}, volume 12337 of {\em LNCS}, pages 251--262, 2020.

\bibitem{BG92}
Richard Beigel and John Gill.
\newblock Counting classes: Thresholds, parity, mods, and fewness.
\newblock {\em Theoretical Computer Science}, 103(1):3--23, 1992.

\bibitem{CH89}
Jin{-}{Y}i Cai and Lane~A. Hemachandra.
\newblock On the power of parity polynomial time.
\newblock In {\em Proc. of the 6th Annual Symposium on Theoretical Aspects of Computer Science, {STACS} 89}, volume 349 of {\em LNCS}, pages 229--239, 1989.

\bibitem{CIK+08}
Chris Calabro, Russell Impagliazzo, Valentine Kabanets, and Ramamohan Paturi.
\newblock The complexity of unique k-{SAT}: An isolation lemma for k-{CNF}s.
\newblock {\em Journal of Computer and System Sciences}, 74(3):386--393, 2008.

\bibitem{Chen92}
Pang~C. Chen.
\newblock Heuristic sampling: A method for predicting the performance of tree searching programs.
\newblock {\em SIAM Journal on Computing}, 21:295--315, 1992.

\bibitem{C15}
Radu Curticapean.
\newblock {\em The simple, little and slow things count : on parameterized counting complexity}.
\newblock PhD thesis, 2015.

\bibitem{FFK94}
Stephen~A. Fenner, Lance Fortnow, and Stuart~A. Kurtz.
\newblock Gap-definable counting classes.
\newblock {\em Journal of Computer and System Sciences}, 48(1):116--148, 1994.

\bibitem{gill}
John Gill.
\newblock Computational complexity of probabilistic {T}uring machines.
\newblock {\em SIAM Journal on Computing}, 6(4):675--695, 1977.

\bibitem{H90}
Ulrich Hertrampf.
\newblock Relations among mod-classes.
\newblock {\em Theoretical Computer Science}, 74(3):325--328, 1990.

\bibitem{KPSZ01}
Aggelos Kiayias, Aris Pagourtzis, Kiron Sharma, and Stathis Zachos.
\newblock Acceptor-definable counting classes.
\newblock In {\em Advances in Informatics, 8th Panhellenic Conference on Informatics, {PCI} 2001. Revised Selected Papers}, volume 2563 of {\em LNCS}, pages 453--463, 2001.

\bibitem{K74}
Donald Knuth.
\newblock Estimating the efficiency of backtrack programs.
\newblock {\em Mathematics of Computation}, 29:122--136, 1974.

\bibitem{KST93}
Johannes K{\"{o}}bler, Uwe Sch{\"{o}}ning, and Jacobo Tor{\'{a}}n.
\newblock {\em The Graph Isomorphism Problem: Its Structural Complexity}.
\newblock Progress in Theoretical Computer Science. Birkh{\"{a}}user/Springer, 1993.

\bibitem{OH93}
Mitsunori Ogiwara and Lane~A. Hemachandra.
\newblock A complexity theory for feasible closure properties.
\newblock {\em Journal of Computer and System Sciences}, 46(3):295--325, 1993.

\bibitem{PZ06}
Aris Pagourtzis and Stathis Zachos.
\newblock The complexity of counting functions with easy decision version.
\newblock In {\em Proc. of the 31st International Symposium on Mathematical Foundations of Computer Science 2006, MFCS 2006}, pages 741--752, 2006.

\bibitem{PZ83}
Christos~H. Papadimitriou and Stathis Zachos.
\newblock Two remarks on the power of counting.
\newblock In {\em Proc. of 6th GI-Conference on Theoretical Computer Science}, volume 145 of {\em LNCS}, pages 269--276, 1983.

\bibitem{PC18}
Tayfun Pay and James~L. Cox.
\newblock An overview of some semantic and syntactic complexity classes.
\newblock {\em Electronic Colloquium on Computational Complexity}, {TR18-166}, 2018.

\bibitem{Purdom78}
Paul~W. Purdom.
\newblock Tree size by partial backtracking.
\newblock {\em SIAM Journal on Computing}, 7(4):481--491, 1978.

\bibitem{RRW94}
Rajesh P.~N. Rao, J{\"{o}}rg Rothe, and Osamu Watanabe.
\newblock Upward separation for {FewP} and related classes.
\newblock {\em Information Processing Letters}, 52(4):175--180, 1994.

\bibitem{Schoning88}
Uwe Sch{\"{o}}ning.
\newblock Graph isomorphism is in the low hierarchy.
\newblock {\em Journal of Computer and System Sciences}, 37(3):312--323, 1988.

\bibitem{Simon75}
Janos Simon.
\newblock {\em On some central problems in computational complexity.}
\newblock PhD thesis, 1975.

\bibitem{Toda91}
Seinosuke Toda.
\newblock {PP} is as hard as the {P}olynomial-{T}ime {H}ierarchy.
\newblock {\em SIAM Journal on Computing}, 20(5):865--877, 1991.

\bibitem{Valiant76}
Leslie~G. Valiant.
\newblock Relative complexity of checking and evaluating.
\newblock {\em Information Processing Letters}, 5(1):20--23, 1976.

\bibitem{Valiant79}
Leslie~G. Valiant.
\newblock The complexity of computing the permanent.
\newblock {\em Theoretical Computer Science}, 8(2):189--201, 1979.

\bibitem{Valiant06}
Leslie~G. Valiant.
\newblock Accidental algorithms.
\newblock In {\em Proc. of the 47th Annual {IEEE} Symposium on Foundations of Computer Science, {FOCS} 2006}, pages 509--517, 2006.

\bibitem{VV86}
Leslie~G. Valiant and Vijay~V. Vazirani.
\newblock {NP} is as easy as detecting unique solutions.
\newblock {\em Theoretical Computer Science}, 47:85--93, 1986.

\end{thebibliography}

\end{document}